\documentclass[journal]{IEEEtran}
\usepackage{xcolor,soul,framed} 
\usepackage{cite}
\usepackage{amsmath,amssymb,amsfonts}
\usepackage{gensymb}
\usepackage{graphicx}
\usepackage{textcomp}
\usepackage{dcolumn}
\usepackage{bm}
\usepackage{subfigure} 
\usepackage{pict2e}
\usepackage{epstopdf}
\usepackage{arydshln}

\newcommand{\Fig}[1]{Fig.\,{\ref{#1}}}

\newcommand{\E}{\mathrm{e}}
\newcommand{\jj}{\mathrm{j}}

\begin{document}

\onecolumn
\newpage
\thispagestyle{empty}

\textbf{Copyright:}
\copyright 2021 IEEE. Personal use of this material is permitted.  Permission from IEEE must be obtained for all other uses, in any current or future media, including reprinting/republishing this material for advertising or promotional purposes, creating new collective works, for resale or redistribution to servers or lists, or reuse of any copyrighted component of this work in other works.\\

\textbf{Disclaimer:} This work has been published in \textit{IEEE Transactions on Antennas and Propagation}. \\

\newpage

\twocolumn

\setcounter{page}{1}

\title{Dispersion Analysis of Periodic  Structures in Anisotropic Media: Application to Liquid Crystals}

\author{Antonio Alex-Amor, Ángel Palomares-Caballero, Francisco Mesa, \textit{Fellow, IEEE}, Oscar Quevedo-Teruel, \textit{Senior Member, IEEE},  and Pablo Padilla. 
\thanks{The work of F. Mesa has been funded by the Consejeria de Transformación Económica, Industria, Conocimiento y Universidades, Spanish Junta de Andalucía (TIC-112) and by the Spanish AEI (Agencia Estatal de Investigación), Ministerio de Ciencia e Innovación (project TEC2017-84724-P) This work was also supported by the Spanish Research and Development National Program under Projects TIN2016-75097-P, RTI2018-102002-A-I00, B-TIC-402-UGR18 and the predoctoral grant FPU18/01965; by Junta de Andalucía under project P18-RT-4830; and by the Spanish Government under Salvador de Madariaga fellowship PRX19/00025.}
\thanks{A. Alex-Amor, A. Palomares-Caballero, and P. Padilla  are with the Departamento de Teor\'{i}a de la Se\~{n}al, Telem\'{a}tica y Comunicaciones, Universidad de Granada, 18071 Granada, Spain (email: aalex@ugr.es, angelpc@ugr.es; pablopadilla@ugr.es.}
\thanks{A. Alex-Amor is also with the Department of Information Technologies, Universidad CEU San Pablo 28003 Madrid, Spain}
\thanks{F. Mesa is with Department of Applied Physics~1,
Escuela Técnica Superior de Ingenieria Informatica, Universidad de Sevilla,
41012 Sevilla, Spain;  (e-mail: mesa@us.es)}
\thanks{O. Quevedo-Teruel is with the Division for Electromagnetic Engineering, School of Electrical Engineering and Computer Science, KTH Royal Institute of Technology, SE-100 44 Stockholm, Sweden (e-mail: oscarqt@kth.se)}
}

\markboth{}%
{Alex-Amor \MakeLowercase{\textit{et al.}}: Dispersion Analysis of Periodic Structures in Anisotropic Media: Application to Liquid Crystals}

\maketitle

\newcommand*{\bigs}[1]{\vcenter{\hbox{\scalebox{2}[8.2]{\ensuremath#1}}}}

\newcommand*{\bigstwo}[1]{\vcenter{\hbox{ \scalebox{1}[4.4]{\ensuremath#1}}}}

\begin{abstract}
This paper presents an efficient method to compute the dispersion diagram of periodic and uniform structures with generic anisotropic media. The method takes advantage of the ability of full-wave commercial simulators to deal with finite structures having anisotropic media. In particular, the proposed method {¡extends the possibilities of commercial eigenmode solvers in the following ways:}¡ (i)~anisotropic materials with non-diagonal permittivity and permeability tensors can be analyzed; (ii)~the attenuation constant can easily be computed in both propagating and stopband regions and lossy materials can be included in the simulation; and~(iii) unbounded and radiating structures such as leaky-wave antennas can be treated. The latter feature may be considered the most remarkable, since the structures must be forcefully bounded with electric/magnetic walls in the eigensolvers of most commercial simulators. In this work, the proposed method is particularized for the study of liquid crystals (LCs) in microwave and antenna devices. Thus, the dispersion properties of a great variety of LC-based configurations are analyzed, from canonical structures, such as waveguide and microstrip, to complex reconfigurable phase shifters in ridge gap-waveguide technology and leaky-wave antennas. Our results have been validated with previously reported works in the literature and with commercial software \textit{CST} and \textit{HFSS}.
\end{abstract}


\IEEEpeerreviewmaketitle

\section{Introduction}

\IEEEPARstart{P}{eriodic} structures are commonly used in many fields of science and engineering. By modifying the geometrical parameters of the unit cell, the propagation of electromagnetic waves throughout the structure can easily be tailored. The addition of tunable materials such as graphene~\cite{graphene1}, ferroelectrics~\cite{ferroelectrics1} or liquid crystal~\cite{lc1} brings an extra degree of reconfigurability to periodic structures. As an example, the radiation properties of antennas \cite{graphene_rad, ferroelectrics_rad, lc_rad} and the phase response of guiding structures \cite{phase_shifter1, phase_shifter2, phase_shifter3} can be electronically controlled, as usually demanded to fulfill the technological challenges of last generation communication systems \cite{5G}. 

The dispersion diagram is the usual scenario to analyze the wave propagation in periodic structures \cite{dispersion_diagram1, Collin}. It gives useful information on the phase velocity, attenuation, radiation losses, coupling between high-order modes, etc. Unfortunately, the anisotropic behavior of the vast majority of tunable materials hampers the computation of dispersion diagrams by general-purpose commercial simulators, even for lossless scenarios. Furthermore, the complex nature of the propagation constant (real and imaginary parts) in lossy and/or radiating structures brings an extra difficulty. In this paper, we propose the use of a multi-modal transfer-matrix method to overcome these weaknesses of the frequency-domain eigenmode solvers of common commercial simulators. The proposed methodology is based on the computation of the general transfer matrix and the resolution of an eigenvalue problem derived from a Floquet analysis \cite{Tsuji1983, Amari2000, Liu2006, Bongard2009, Marini2010, Islam2011, Coves2012, Weitsch2012, Naqui2012, Collin, symmetry_multimode, bloch_mtt, multimodal_apm}. In particular, the use of the proposed multi-modal approach offers three main advantages over the eigensolver tools of commercial simulators when analyzing periodic structures with generic anisotropic media:

\begin{enumerate}
    \item 
    Anisotropic materials with non-diagonal permittivity and permeability tensors can be considered. It should be noted that most commercial simulators normally deal with anisotropic materials with diagonal tensors. However, for some particular configurations involving non-diagonal tensors, commercial simulators find difficulties.
    \item 
    The attenuation constant can easily be computed with the multi-modal method in both propagating and stopband regions. Furthermore, lossy materials can be included in the computation. This will be exploited in Sec.\,\ref{sec:Shifter} for the  analysis of a reconfigurable phase shifter based on a liquid crystal.
    \item 
    Unbounded and radiating structures can be treated, unlike what happens in most commercial simulators where the structure has to be bounded with perfect electric/magnetic conditions. This is a remarkable feature that will be exploited in Sec.\,\ref{sec:LWA} for the analysis and optimization of a reconfigurable leaky-wave antenna that uses a liquid crystal as a tunable material.
\end{enumerate}

Liquid crystal is one of the most promising tunable materials for applications in the microwave range~\cite{crystals}. However, the study of the wave propagation in liquid-crystal-based periodic structures is a complex task due to the anisotropic and lossy nature of the material, accounted for by a non-diagonal permittivity tensor~\cite{lc_book1}. As a consequence, the multi-modal transfer-matrix  method  arises  as  an  interesting  option  to  analyze periodic  structures  with  liquid  crystal  material in the design of electronically reconfigurable devices.

The paper is organized as follows. Sec.\,\ref{sec:Theory} describes the formulation of the multi-modal transfer-matrix method for periodic structures in general anisotropic media. Then, we particularize to the use of liquid crystal and the main properties of this material are summarized. In Sec.\,\ref{sec:WG} the dispersion properties of rectangular and parallel-plate waveguides filled with liquid crystal are analyzed. Sec.\,\ref{sec:MS} analyzes the dispersion properties of microstrip lines suspended on liquid crystal substrates. The analysis of the dispersion properties of an electrically reconfigurable phase shifter in ridge gap-waveguide technology is carried out in~Sec.\ref{sec:Shifter}. In Sec.\,\ref{sec:LWA} the design of a electrically reconfigurable leaky-wave antenna is discussed. It should be remarked that the results presented in Secs.\,\ref{sec:WG}-\ref{sec:LWA} have been validated with previously reported data in the literature and with the Eigenmode solver of~\textit{CST}. Finally, the main conclusions of the work are drawn in Sec.\,\ref{sec:Conc}.

\section{\label{sec:Theory} Theoretical Framework}

\subsection{\label{sec:MM} Multi-modal Analysis}

The Multi-Modal Transfer-Matrix Method (MMTMM) applied to the computation of periodic structures in anisotropic media is briefly outlined in this section. For the sake of simplicity, we focus on the study of 1-D periodic structures, although the analysis can be straightforwardly extended to 2-D periodic structures \cite{bloch_mtt, multimodal_apm}. 

For a 1-D periodic structure, the eigenvalue problem that leads to the dispersion relation is \cite{Collin}
\begin{equation} \label{eigenproblem1}
    [\mathbf{T}] \begin{bmatrix} \mathbf{V} \\ \mathbf{I} \end{bmatrix} = \E^{\gamma p} \begin{bmatrix} \mathbf{V} \\ \mathbf{I} \end{bmatrix}
\end{equation}
where $[\textbf{T}]$ is the $2N \times 2N$ multi-modal transfer matrix, with $N$~being the number of modes considered in the computation, \textbf{V} and~\textbf{I} are $N \times 1$ arrays containing the voltages and currents at the output ports, $\gamma=\alpha + \mathrm{j}\beta$ is the propagation constant, $\alpha$ is the attenuation constant, $\beta$ is the phase constant,  and $p$~is the period of the unit cell. The transfer matrix, which is partitioned in four $N \times N$ submatrices $[\mathbf{A}]$, $[\mathbf{B}]$, $[\mathbf{C}]$, and~$[\mathbf{D}]$ as
can be derived from the generalized multi-mode scattering matrix $[\textbf{S}]$, as detailed in \cite{matrices1, matrices2} or, alternatively, by means of the algebraic manipulations presented, for instance, in~\cite[Eq.\,(1)]{Coves2012}.

In the case that the structure under study is symmetrical and reciprocal ($[\mathbf{A}]=[\mathbf{D}]^H$, $[\mathbf{B}]=[\mathbf{B}]^H$, $[\mathbf{C}]=[\mathbf{C}]^H$), the original $2N$-rank eigenvalue problem in \eqref{eigenproblem1} can be simplified to the $N$-rank eigenvalue problem \cite{multimodal_apm}
\begin{equation} \label{eigenproblem2}
    [\mathbf{A}]\mathbf{V} =\cosh (\gamma p)\mathbf{V}\;.
\end{equation}
%

The multi-modal scattering matrix $[\mathbf{S}]$ of a periodic structure can be computed via full-wave simulations of a single unit cell. Inter-cell coupling  effects are taken into account through the higher-order modes used in the multi-mode representation~\cite{symmetry_multimode}. Commercial simulators or in-house codes can be utilized for this purpose. In this work, we make use of the commercial software \textit{CST} and \textit{Ansys \textit{HFSS}} for the extraction of the scattering matrix~$[\mathbf{S}]$. It should be remarked that the scattering parameters of finite structures can be computed in anisotropic media (including losses and non-diagonal tensorial materials) with the time-domain and frequency-domain solvers of \textit{CST}.  However, the dispersion diagrams of the periodic structure cannot directly be  computed with the \textit{CST} Eigenmode solver, unless the anisotropic material is lossless and defined by a diagonal tensor. Therefore, the proposed hybrid implementation benefits from the use of commercial simulators to obtain the multi-modal transfer matrix and then compute the dispersion properties of periodic structures in anisotropic media by solving the corresponding eigenvalue problem.

\subsection{\label{sec:LC} Liquid Crystal}

\begin{figure}[t]
	\centering
	\subfigure[]{\includegraphics[width= 0.4\columnwidth]{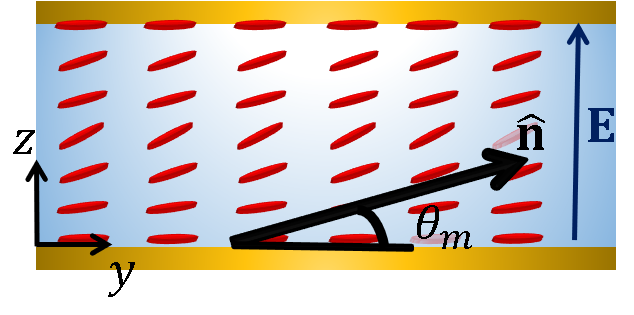}
	}\qquad 
	\subfigure[]{\includegraphics[width= 0.4\columnwidth]{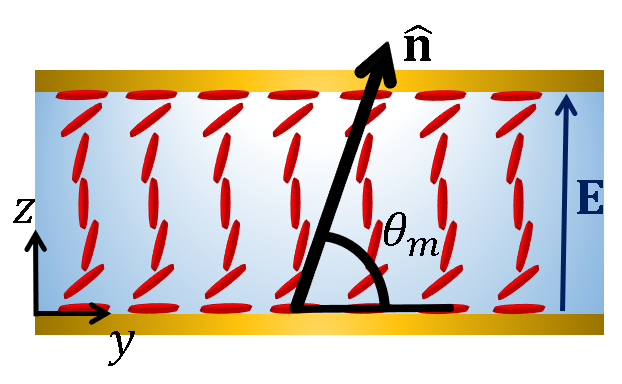}
	}
	\caption{Molecular reorientation in a nematic liquid crystal enclosed by two metallic plates when a (a) low-intensity and (b) high-intensity quasi-static electric field $\mathbf{E}$ is applied. Unit vector $\mathbf{\hat{n}}$ represents the average director and $\theta_m$ the average tilt angle.} 
	\label{lc_director}
\end{figure}

As is well known, liquid crystal (LC) is a state of matter that combines properties of liquids and solid crystals. The elongated rod-like shape of molecules in LCs gives the material its characteristic anisotropic behavior, defined by the fast and slow propagation axes. Depending on the type of order of the molecules, there exist different states or mesophases in which LCs can be found: nematic, smetic and cholesteric \cite{lc_book1, lc_book2}. From all the mentioned states, nematic LCs have demonstrated to be particularly useful for the design of reconfigurable radio-frequency devices, such as filters \cite{LCfilter1, LCfilter2, LCfilter3}, antennas \cite{lc_rad, gerardoLC_layers, LCantenna2, LCantenna3}, frequency selective surfaces \cite{FSSLC2008, maciLC2017} and phase shifters \cite{phase_shifter3, ridgegap_phaseshifter, lc_phaseshifter3}. In a nematic LC, molecules are oriented in the same average direction, represented by the director $\mathbf{\hat{n}}$ and the average tilt angle $\theta_m$. Molecules can be reoriented with the use of magnetic or electric fields \cite{lc_book1, lc_book2}. If LCs are enclosed between metallic plates, which could be the case of waveguides, parallel plates and microstrip lines (see~\Fig{lc_director}), quasi-static electric fields are normally used for simplicity to polarize the material. Molecules tend to orient parallel to the metallic plates when a low-intensity electric field is applied [\Fig{lc_director}(a)] whereas molecules orient perpendicular to the metallic plates when a high-intensity electric field is applied [\Fig{lc_director}(b)]. 

The uniaxial permittivity tensor that characterizes the electrical properties of the LC can be expressed as \cite{lc1, lc_book1}
\begin{equation} \label{tensor}
    \overline{\varepsilon}=
    \begin{pmatrix}
        \varepsilon_\bot & 0 & 0\\
        0 & \varepsilon_\bot+ \Delta \varepsilon \cos^2\theta_m & \Delta\varepsilon \cos \theta_m \sin \theta_m \\
        0 & \Delta\varepsilon \cos \theta_m \sin \theta_m & \varepsilon_\bot+ \Delta \varepsilon \sin^2\theta_m
    \end{pmatrix}
\end{equation}
where $\Delta\varepsilon=\varepsilon_{\parallel}-\varepsilon_\bot$ is the dielectric anisotropy, $\varepsilon_{\parallel}$ is the parallel permittivity and $\varepsilon_\bot$ is the perpendicular permittivity. The loss tangent tensor is calculated analogously, by replacing $\Delta \varepsilon$ and $\varepsilon_\bot$ in~\eqref{tensor} by the anisotropic loss tangent $\Delta\tan \delta=\tan \delta_{\parallel}-\tan \delta_{\bot}$ and the perpendicular loss tangent $\tan \delta_{\bot}$, respectively. As previously stated, $\theta_m$ represents the average tilt angle of the molecules. This angle is a function of the elastic constants $k_{ii}$ \cite{elastic_constants}, the dielectric anisotropy at the bias frequency $\Delta\varepsilon^b$, the intensity of the quasi-static electric field  and the pretilt angle $\theta_p \simeq 0$. That is, $\theta_m=\theta_m(k_{11}, k_{22}, k_{33}, \Delta \varepsilon^b, V, \theta_p)$. A detailed study particularized to LCs enclosed in parallel-plate waveguides, showing the relation between $\theta_m$ and the parameters presented above, can be found in \cite{antonio_lc1, antonio_lc2}.

The permittivity tensor \eqref{tensor} becomes a diagonal tensor for the extreme cases of $\theta_m=0^\mathrm{o}$ and $\theta_m=90^\mathrm{o}$, when the LC is polarized with $V=0\,$V (with $\theta_p=0$), and an hypothetical infinite voltage $V_\infty$, respectively. As previously discussed, eigensolver tools of commercial simulators find difficulties when computing the dispersion diagram of periodic structures composed by anisotropic materials that are defined by non-diagonal permittivity and permeability tensors; that is, whenever $\theta_m \neq 0,90^\mathrm{o}$ in the case of working with liquid crystals.
Furthermore, losses can easily be included with the proposed method. 

\subsection{Practical Considerations}

As detailed in Sec.\,\ref{sec:MM}, the scattering parameters can be computed via full-wave simulations with either in-house codes or commercial simulators. In this work, we take advantage of the ability of the commercial simulators \textit{CST} and \textit{Ansys HFSS} to deal with arbitrary geometries and materials. In these simulators, two different solvers are offered to compute the scattering parameters of a single unit cell: time-domain solver and frequency-domain solver. Some relevant characteristics of these solvers are detailed next (these characteristics are referred to version \textit{CST}~2020 and \textit{HFSS}~18, and can differ for other software and/or versions).

(1)\textit{ Time-domain solvers}: \textit{CST} can only deal with diagonal tensors in the simulation, only the extreme polarization states $\theta_m=0,90$º can be computed (see Sec.\,\ref{sec:LC}).  Losses can be included in the computation of the scattering parameters. Input and output ports can be directly placed at the surface of the anisotropic material (i.e., liquid crystal in our case). The Transient solver of \textit{HFSS} is not able to analyze anisotropic materials.
    
(2)\textit{ Frequency-domain solvers}: \textit{CST} can deal with non-diagonal permittivity/permeability tensors via the macro ``Full Tensor Material''. Losses can be included in the computation. However, input and output ports cannot be directly placed at the surface of the anisotropic material. It means that isotropic layers have to be added to feed the structure, which requires a de-embedding process to characterize the structure under study. In \textit{HFSS}, when a driven modal solution is set, it is also able to perform simulations with anisotropic materials including losses. The de-embedding process is not required as long as the tensor is diagonal.

In this work, as long as only extreme polarization states ($\theta_m=0,90\degree$) need to be simulated, we have opted to use the time-domain solver of \textit{CST} because the de-embedding of the ports can be avoided. Both \textit{HFSS} and \textit{CST} frequency-domain solvers offer a greater versatility, as they allow for dealing with non-diagonal tensors. However, in the case of using \textit{CST}, it is required to perform a de-embedding of the input and output ports.

In general, the correct choice of the input/output modes in the unit cell of the periodic structure plays a fundamental role when applying the MMTMM. In this choice, it should be considered that some of the modes which are excited at the input/output ports do not have a physical correspondence with the Bloch modes that actually propagate in the unit cell. As the virtual waveguides associated with the input/output ports are constrained by the boundary conditions imposed to these ports, the modes of these virtual waveguides are often hardly related to the actual modes of the periodic structure.  In order to ensure a correct convergence of the method, these irrelevant  modes should not be considered~\cite{multimodal_apm}.

In most practical cases, the exciting ports inside the commercial simulator cannot be directly placed at the interface of an anisotropic material. For that reason, we are usually forced to add two de-embedding (isotropic) layers at the input and output sections of the structure. Fortunately, most commercial software offer the possibility to change the reference planes of the input and output ports. Geometrical and electrical parameters of the de-embedding blocks should be carefully chosen in order to  minimize the error and avoid unwanted resonances. First, it is recommended to set the length of the de-embedding layers as short as possible. Secondly, their relative permittivity and permeability  should be selected to match the average (effective) value of the tensors that model the electromagnetic behavior of the considered anisotropic material. In the case of liquid crystals, the relative permeability of the de-embedding layers is set to~1 and the relative permittivity  to 
\begin{equation}
    \varepsilon_\mathrm{eff}=\sqrt{(\varepsilon_{\parallel} \sin \theta_m)^2 + (\varepsilon_{\bot} \cos \theta_m)^2 }\;.
\end{equation}
Slight variations of the mentioned values could apply for a finer adjustment.

Concerning the case of \textit{open and/or radiating structures}, both \textit{HFSS} and \textit{CST} simulators intrinsically bound the input and output ports with electric or magnetic walls in order to properly excite the considered structure.  Of course, these boundary conditions lead to a discrete spectrum and, subsequently, there is always an intrinsic approximation in the estimation of the spectrum of open (radiating) structures, as detailed in \cite{multimodal_apm}. Nonetheless, if the port dimensions are correctly selected, it is found that the MMTMM can accurately compute the dispersion properties in a wide variety of practical designs, such as the LC-based leaky-wave antenna presented in Sec. VI.  

A good selection of the port should ensure that most of the radiated energy (or field intensity) is allocated within the enclosure of the input and output ports. Under this basic assumption, we can take advantage of the \textit{a priori} physical knowledge about the  modes  that  can  propagate  in  the  structure to set the port dimensions. As an example, the excited TEM mode in a parallel-plate structure with a small plate aperture is expected to be confined in the area between the metallic plates.  However, the field intensity in microstrip and coplanar structures is less confined around the main line(s),  so the port size should be larger compared to a parallel-plate configuration. Once the input/output ports are set, it is advisable to check that further increase in the port dimensions does  not  affect  the  results  of  the  scattering  parameters. 

\medskip

\section{\label{sec:WG} Metallic Waveguide}

The study of metallic waveguides filled with skew uniaxial dielectrics has been a topic of continuous interest in microwave/antennas engineering~\cite{Berk56, Davies67, Sun16}. It is well known that, in general, these waveguides support hybrid modes, and only in some particular cases there exist pure TE/TM modes in the waveguide. In our LC case under study, given the structure of the permittivity tensor in~\eqref{tensor}, the modes of the LC-filled waveguide will be hybrid when $\theta_m\neq 0,90$º, thus requiring the MMTMM to obtain accurate solutions. Here it should be noted that, although the MMTMM is originally posed to deal with periodic structures, it can also be applied to the computation of the dispersion diagram of uniform (non-periodic) structures. This computation is found sufficiently accurate provided that the phase shift, $\beta L$, is not close to the edges of the first Brillouin zone, where~$L$ is the length of the considered waveguide section. It is then advisable to use small values of~$L$ when computing the scattering parameters of the waveguide in the commercial simulator. 

\begin{table}[t]
\centering
\caption{
Guide wavelength, $\lambda_g$, as a function of the free-space wavelength, $\lambda_0$, both in mm, for the test case presented in~\cite{Davies67}}
\label{table_davies}
\begin{tabular}{ccccc}
\hline
\textbf{$\lambda_0$} & \textbf{$N=1$} & \textbf{$N=2$} & \textbf{$N=3$} & \cite{Davies67} \\ \hline\hline
7   & 0.6512    & 0.5886   & 0.5878    & 0.5852    \\ \hline
8.5 & 0.8525    & 0.7953   & 0.7948     & 0.7932    \\ \hline
10  & 1.1410    & 1.0787   &  1.0756    & 1.0680    \\ \hline
12  & 1.8487    & 1.7782   & 1.7573    & 1.6496    \\ \hline
15  & $\jj 3.6047$ & $\jj 4.9223$ & $\jj 4.9478$ & $\jj 5.0090$ \\ \hline
\end{tabular}
\end{table}

Before studying the LC-filled waveguides, a validation of the method will be carried out by comparing the results reported in~\cite[Table~I]{Davies67} for a metallic waveguide filled with a strongly anisotropic skew uniaxial dielectric with the ones provided by the MMTMM. This comparison is shown in~Table\,\ref{table_davies}, where our results with $N=1$ means that only the TE$_{10}$ is employed in the input/output port, $N=2$ stands for an additional TE$_{01}$ mode, and $N=3$ for an additional TE$_{21}$ mode.  This consecutive addition of modes follows the rationale in~\cite{Davies67} for the first, second, and third approximations there discussed. In particular, the column data from~\cite{Davies67} in the table correspond to the third approximation.

\begin{figure}[h]
	\centering
	\subfigure{\includegraphics[width= 0.7\columnwidth]{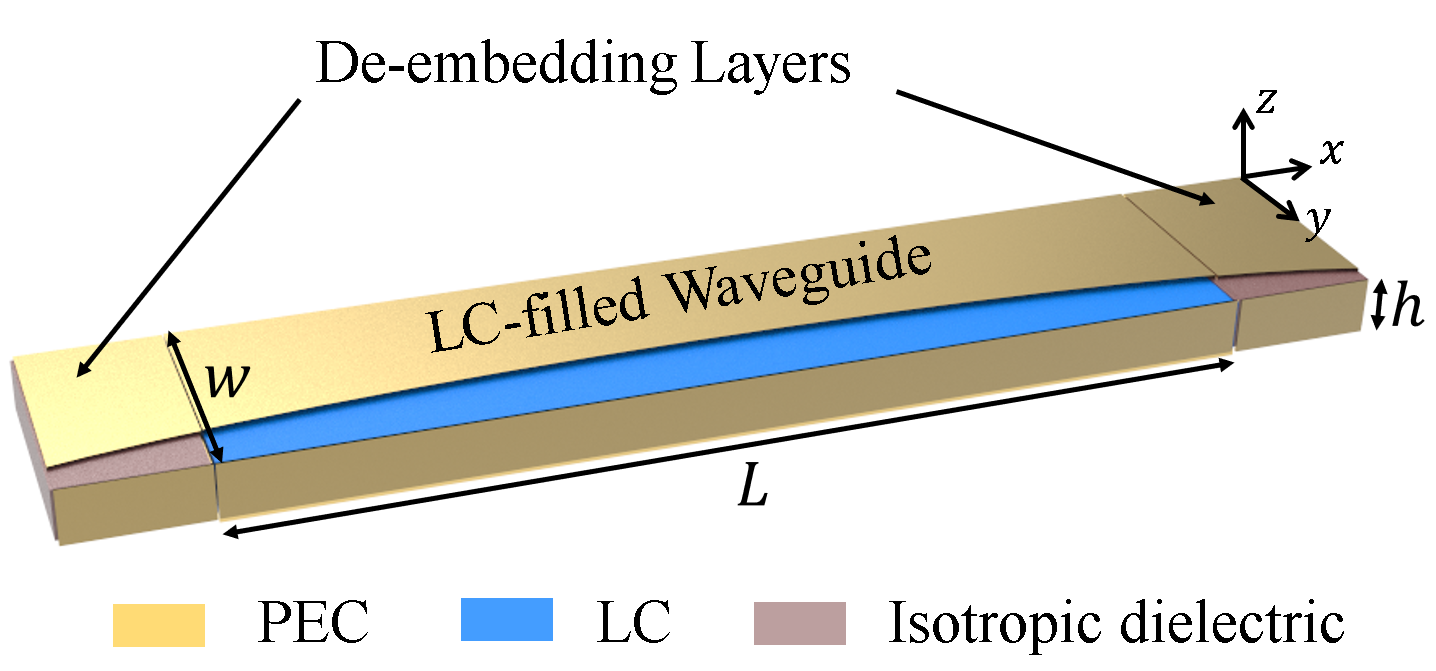}}
	\subfigure{\includegraphics[width= 0.97\columnwidth]{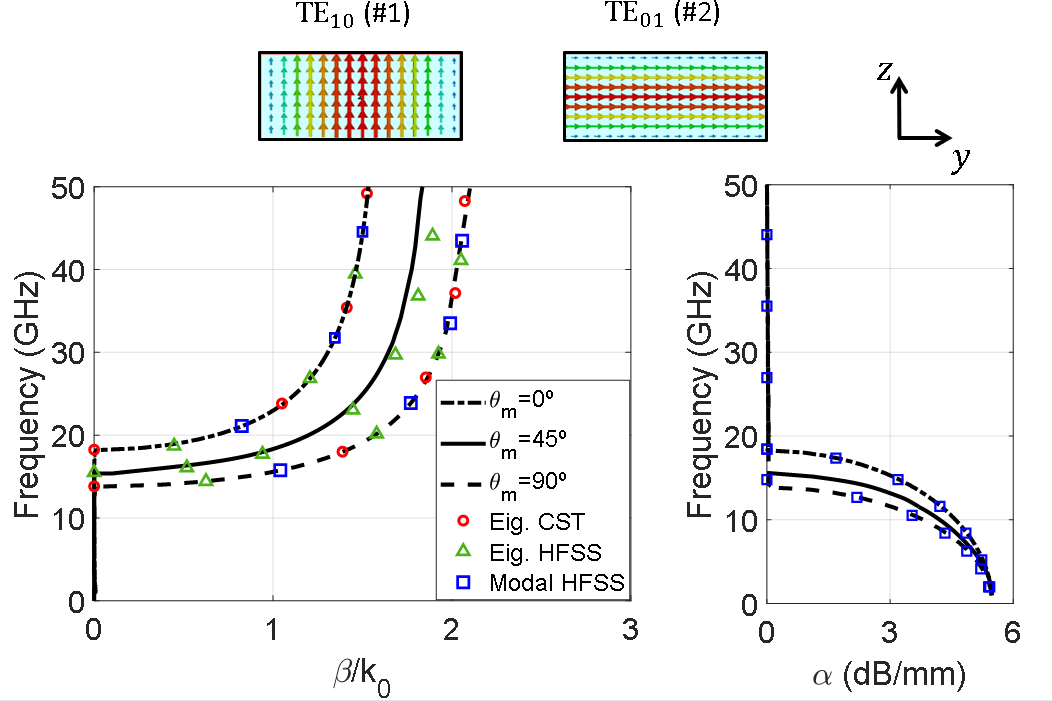}
	}
	\caption{Dispersion diagram of a metallic rectangular waveguide filled with lossless liquid crystal. $N=2$ modes have been used for the computation of the hybrid mode in the case $\theta=45^\mathrm{o}$. Two de-embedding (isotropic) layers are placed to correctly feed the structure. The electrical field distribution at the input and output de-embedding  layers is shown in the figure. The electrical and geometrical parameters of the liquid crystal cell are: $\varepsilon_\bot=2.7$, $\Delta\varepsilon=2$, $L=0.01$\,mm, $w=5$\,mm, $h=2.5$\,mm.}
	\label{section_PEC}
\end{figure}

The computation of the generalized scattering matrix from the commercial simulator has followed the general recommendations given in Section II.C. Thus, two short isotropic layers (see Fig.\,\ref{section_PEC}) have been placed at the input and output sections in order to properly feed the anisotropic waveguide. The whole structure has been simulated with the frequency-domain solver in \textit{CST} and \textit{Ansys \textit{HFSS}}. Our results show a good convergence as~$N$ increases as well as a good agreement with those from~\cite{Davies67}. As commented in~\cite{Davies67}, the skew anisotropy actually requires the hybrid modal solutions of the waveguide to be constructed in general with multiple TE and TM modes (the required number to achieve accuracy will depend on the operation frequency). However, in situations where the off-diagonal elements of the permittivity tensors are smaller than the diagonal ones, the number of required modes could be reduced to a pair of modes or even just a single mode.

Once the MMTMM has been conveniently validated, it will be used to analyze the dispersion properties of a uniform metallic waveguide (along the $x$-direction) filled with a LC. \Fig{section_PEC} shows the dispersion diagram of both the  phase ($\beta$) and attenuation ($\alpha$) constants for different values of the average orientation of the molecules in the LC. When $\theta_m=0,90$º, the permittivity tensor is diagonal and, according to \cite{Davies67,Sun16}, pure TE and TM modes exist. In particular, when the optic axis is oriented along the $y$ axis ($\theta_m=0$º), the fundamental mode is the TM$_{Y10}$, following the notation of~\cite{Sun16}. When the optic axis is oriented along the $z$ axis ($\theta_m=90$º), the fundamental mode is the TE$_{Z10}$. Both TM$_{Y10}$ and TE$_{Z10}$ modes have correspondence with the TE$_{10}$ mode in a conventional waveguide loaded with an homogeneous isotropic dielectric~\cite{Sun16}. This means that only one mode ($N=1$) is required in the MMTMM to obtain accurate results in both $\theta_m=0,90$º cases. 
In fact, an excellent agreement is observed in~\Fig{section_PEC} between the proposed method and commercial software \textit{CST} and \textit{HFSS}
for these cases. Only the phase constant of these two extreme cases can be computed with the eigenmode solver of \textit{CST}, corresponding to the situations in which the tensor is diagonal and no losses are considered. Furthermore, the eigenmode solver of \textit{CST} cannot compute the attenuation constant in periodic structures, not even in lossless and bounded isotropic scenarios.  On the other hand, the eigensolver of \textit{HFSS} is able to include losses in the computation and deal with non-diagonal tensors. By post-processing the so-called ``complex frequencies", the attenuation constant can be extracted from this software. However, the eigensolver of \textit{HFSS} cannot compute the complex frequencies (and therefore, neither the attenuation constant) in the stopband regions of periodic structures and in the cutoff region of a waveguide, as detailed in  \cite{complexfrequencies}. For that reason, when $\theta_m\neq0,90$º, no comparison with \textit{HFSS} results is provided for the attenuation constant $\alpha$ in \Fig{section_PEC}. Nonetheless, the Driven Modal solver of \textit{HFSS} can be efficiently used to compute the complex propagation constant in uniform structures, such as the LC-filled waveguide shown  in \Fig{section_PEC}, as long as the permittivity and permeability tensors are expressed in diagonal form. Good agreement is observed between the driven modal solution of \textit{HFSS} and our MMTMM. 

The proposed multi-modal approach can help us in these situations, providing accurate results for both phase and attenuation constants in all polarization states of the LC. In the case shown in~\Fig{section_PEC} for $\theta_m=45^\mathrm{o}$, $N=2$ modes in the input/output ports suffice to achieve convergence in the considered frequency range; namely, the TE$_{10}$ and the TE$_{01}$ modes at the isotropic interfaces. Good agreement is observed between the MMTMM and the eigensolver of \textit{HFSS} for the computation of the phase constant in the case $\theta_m=45^\mathrm{o}$. Note that \textit{HFSS} cannot compute the attenuation constant in this particular case, since cannot provide the complex frequency in the cutoff region of the waveguide.


\section{\label{sec:MS} Microstrip-like line}
Previous works  have reported complex numerical techniques to study the wave propagation in microstrip lines filled with liquid crystal~\cite{microstripLC_MTT, microstripLC_congress}. In this section, we show that the dispersion properties of a LC-based microstrip lines can be analyzed in an alternative and easier manner with the use of the MMTMM. Specifically, the structure analyzed in~\cite{microstripLC_MTT} is now analyzed with the MMTMM and the results compared with measurement and numerical (Finite Elements Method, FEM) data provided in~\cite{microstripLC_MTT}. 

\begin{figure}[t]
	\centering
	\subfigure[]{\includegraphics[width= 0.75\columnwidth]{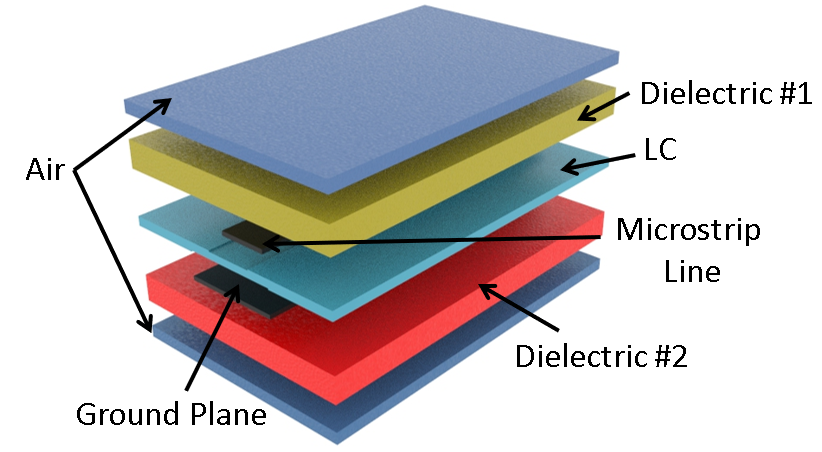}
	}
	\subfigure[]{\includegraphics[width= 0.65\columnwidth]{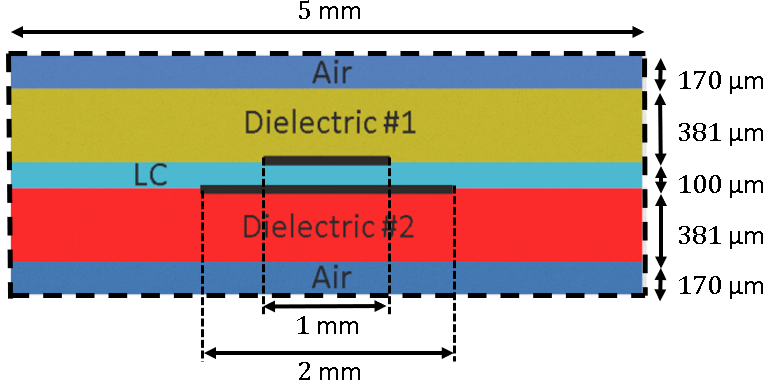}
	}
	\caption{Microstrip section filled with liquid crystal presented in \cite{microstripLC_MTT}: (a) 3-D view, (b) transversal cut.} 
	\label{microstrip}
\end{figure}

 A 3-D view of the microstrip structure under study is depicted in~\Fig{microstrip}(a) with \Fig{microstrip}(b) showing a transversal cut with the different layers and dimensions. A Merck~E7 liquid crystal of electrical parameters $\varepsilon_\bot=2.78$, $\Delta\varepsilon=0.47$, $k_{11}=11.1$ pN, $k_{22}=10.0$ pN, $k_{33}=17.1$ pN, $\theta_p=2^\mathrm{o}$ and  $\Delta\varepsilon^b(1 \, \mathrm{kHz})=13.8$ was chosen in~\cite{microstripLC_MTT}. Note that the elastic constants $k_{ii}$, the pretilt angle $\theta_p$ and the dielectric anisotropy at the bias frequency $\Delta\varepsilon^b$ (see Sec.\,\ref{sec:LC}) are given in this case to relate a determined polarization voltage $V$ with the resulting average tilt angle $\theta_m$. Two dielectrics of permittivity $\varepsilon_{r}^{\mathrm{diel1}}=3.27$, and $\varepsilon_{r}^{\mathrm{diel2}}=9.8$ are used to encapsulate the~LC. For the computation, the length of the microstrip section is $L=1$ mm. As in~\cite{microstripLC_MTT}, the structure is shielded by applying electric boundary conditions  to the edges of~\Fig{microstrip}(b), represented by black dashed lines. 

\begin{figure}[t]
	\centering
	\subfigure{\includegraphics[width= 0.95\columnwidth]{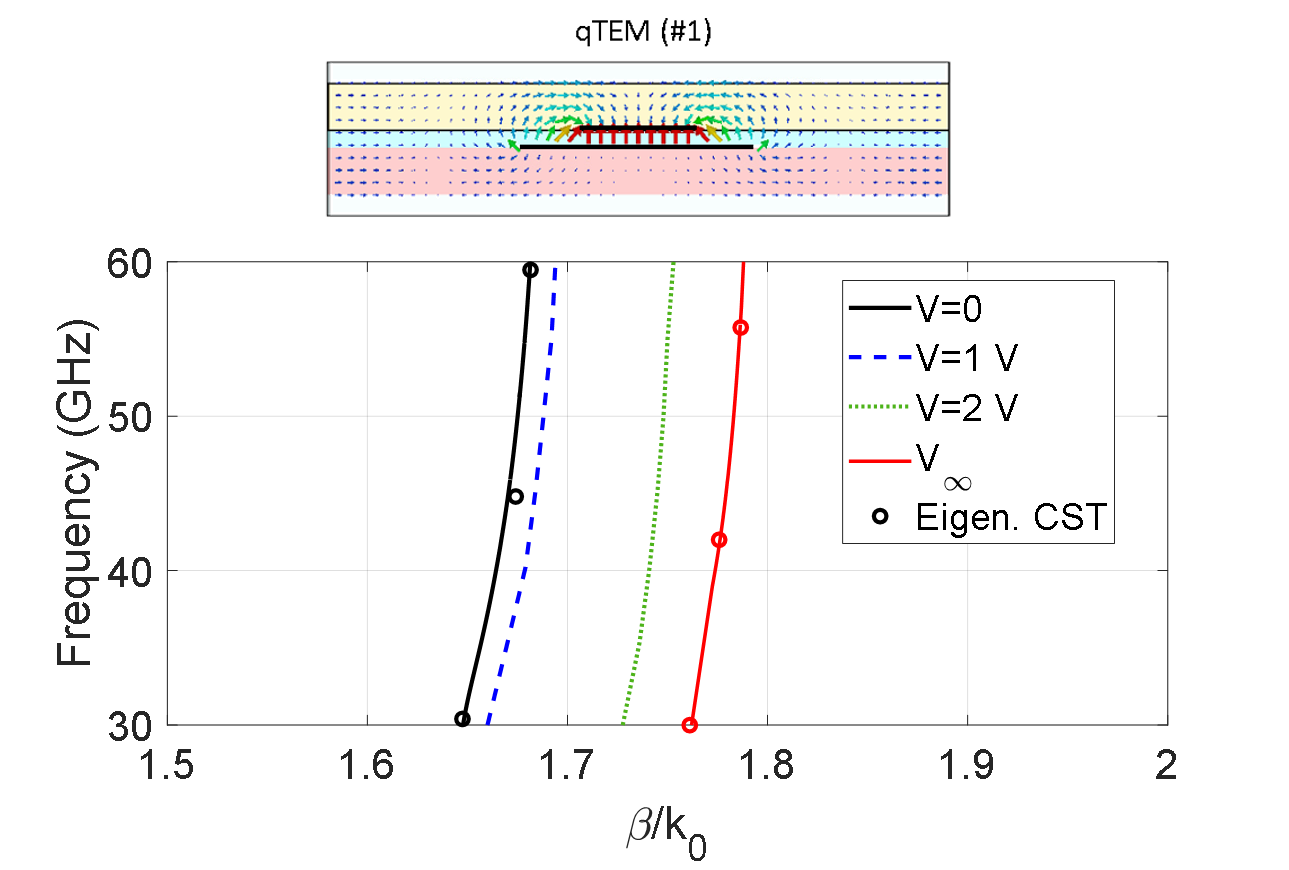}
	}
	\caption{Dispersion diagram of the lossless microstrip-like line with a liquid crystal cell presented in~\cite{microstripLC_MTT} for the extreme and some intermediate polarization stages ($N=1$ mode). The electrical field distribution of the considered mode is shown in the figure. The electrical parameters of the liquid crystal cell and dielectrics are $\varepsilon_\bot=2.78$, $\Delta\varepsilon=0.47$, $k_{11}=11.1$ pN, $k_{22}=10.0$ pN, $k_{33}=17.1$ pN, $\theta_p=2^\mathrm{o}$, $\Delta\varepsilon^b(1 \, \mathrm{kHz})=13.8$, $\theta_p=2^\mathrm{o}$, $\varepsilon_{r}^{\mathrm{diel1}}=3.27$, and $\varepsilon_{r}^{\mathrm{diel2}}=9.8$.} 
	\label{microstrip_dispersion}
\end{figure}

The dispersion diagram of the lossless LC-based microstrip section presented in~\cite{microstripLC_MTT} is shown in~\Fig{microstrip_dispersion} for different polarization voltages. The scattering parameters were extracted from a frequency-domain simulation in commercial software \textit{CST}. In our computations with the MMTMM, it is found that the use of just the fundamental propagating qTEM mode ($N=1$) suffices to provide accurate enough results.  This is confirmed with the good agreement found with the results extracted from the eigenmode solver of~\textit{CST} as well as the obtaining of a negligible attenuation constant ($\alpha/k_0<5\!\times\!10^{-3}$) in all the considered frequency range. Additionally, it should be remarked that a null polarization voltage ($V=0$) does not represent in this particular case a diagonal tensor, since the pretilt angle is different from zero ($\theta_p=2$º) here. Thus, the average tilt angle associated with a null polarization voltage will be approximately $\theta_m \approx \theta_p =2$º, resulting in a non-diagonal tensor. As a consequence, the \textit{CST} Eigenmode solver cannot actually compute the case $V=0$, although it can be approximated with almost negligible error to a diagonal tensor ($\theta_m \approx 0$º).

 In order to form the permittivity tensor~\eqref{tensor} and then compute the S-parameters of the structure in \textit{CST}, a conversion between the polarization voltage~$V$ and the average tilt angle~$\theta_m$ has to be done. This conversion has been carried out with the formulas of \cite{antonio_lc2}, which can also be applied in microstrip structures as long as fringing-field effects are not relevant. Looking at the phase constant in~\Fig{microstrip_dispersion}, it can be appreciated that the structure becomes denser as the polarization voltage increases. Furthermore, the LC is almost saturated (with respect to $V_\infty$) for a very low voltage values such as $V = 2$\,V\, due to the elevated dielectric anisotropy $\Delta\varepsilon^b$ that~Merck~E7~LC possesses at the bias frequency. 
 

\begin{figure}[!t]
	\centering
	\subfigure{\includegraphics[width= 0.9\columnwidth]{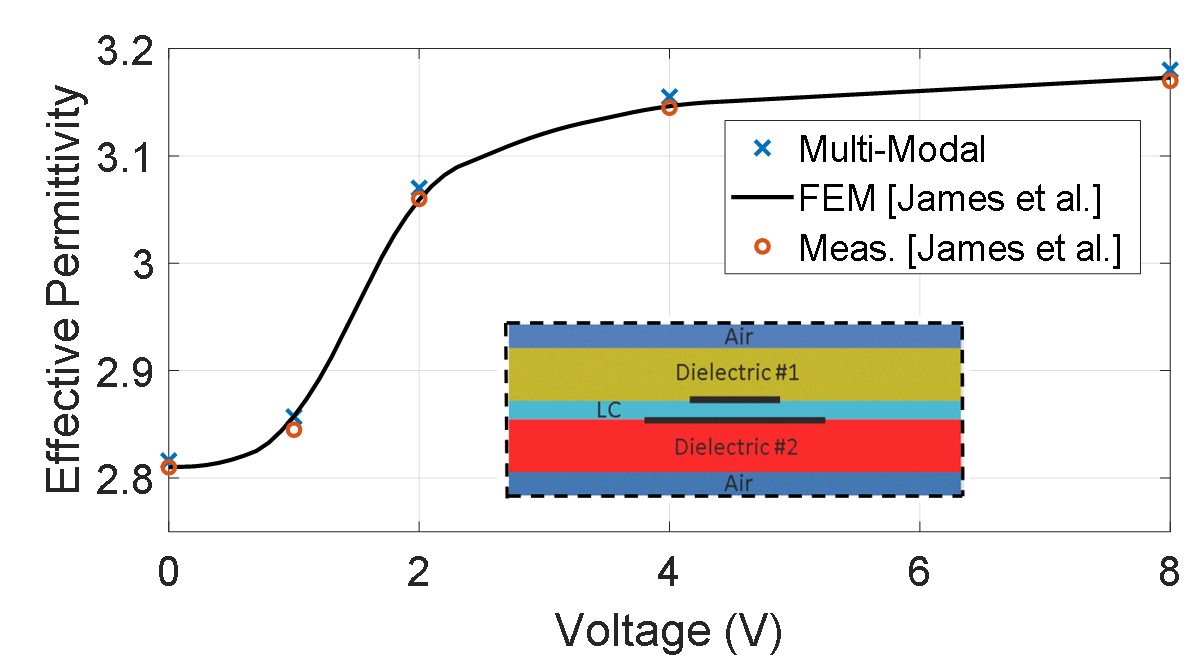}
	}
	\caption{Effective relative permittivity at 60\,GHz for different polarization voltages. A comparison is made with the numerical results and mesurements presented in \cite{microstripLC_MTT}. $N=1$ mode is used for the computation. The electrical parameters of the liquid crystal cell and dielectrics are $\varepsilon_\bot=2.78$, $\Delta\varepsilon=0.47$, $k_{11}=11.1$ pN, $k_{22}=10.0$ pN, $k_{33}=17.1$ pN, $\theta_p=2^\mathrm{o}$, $\Delta\varepsilon^b(1 \, \mathrm{kHz})=13.8$, $\theta_p=2^\mathrm{o}$, $\varepsilon_{r}^\mathrm{diel1}=3.27$, and $\varepsilon_{r}^{\mathrm{diel2}}=9.8$.} 
	\label{microstrip_effective}
\end{figure}


In \Fig{microstrip_effective}, the  relative effective permittivity of the structure at~60\,GHz is shown for different polarization voltages. This permittivity is computed as $\varepsilon_{r,\text{eff}}(f)=\beta^2(f)/k_0^2(f)$, which is directly extracted from the dispersion diagrams in~\Fig{microstrip_dispersion}. An excellent agreement is observed with the FEM and measurement data reported in~\cite{microstripLC_MTT}. As previously discussed, the effective permittivity rapidly saturates for low polarization voltages due to the elevated dielectric anisotropy at the bias frequency. Note that the maximum effective relative permittivity would be approximately $\varepsilon_{r,\text{eff}} \approx \varepsilon_{\parallel}= 3.25$ for $V\rightarrow\infty$, and that $\varepsilon_{r,\text{eff}}(V=8\,\text{V})=3.18$ is already close to this value.

\section{\label{sec:Shifter} Reconfigurable Phase Shifter}

In this section we present the design and analysis of a LC-based reconfigurable phase shifter in ridge gap-waveguide technology. The dispersion properties of the reconfigurable phase shifter in lossless and lossy scenarios are computed by means of the MMTMM. This task could be of potential interest for the development of efficient tunable phase shifters applied to the design of phased array antennas.

\begin{figure}[t]
	\centering
	\subfigure[]{\includegraphics[width= 0.45\columnwidth]{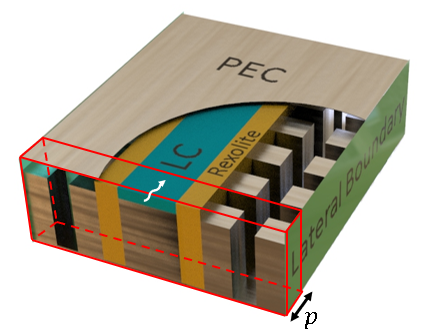}
	} %
	\subfigure[]{\includegraphics[width= 0.5\columnwidth]{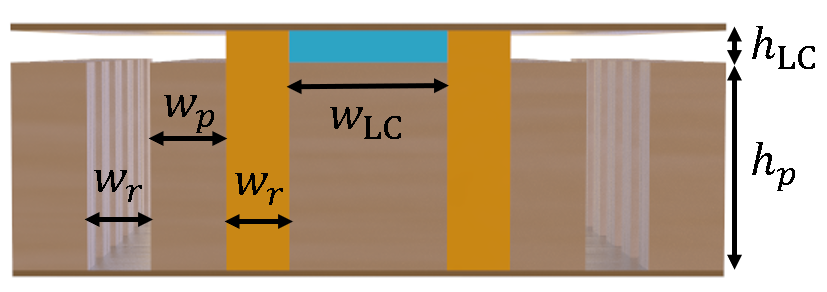}
	}
	\caption{Liquid-crystal-based reconfigurable phase shifter in ridge gap-waveguide technology.} 
	\label{ridge_design}
\end{figure}

An schematic of the phase shifter in ridge gap-waveguide technology and its transversal cut view are shown in Figs.\,\ref{ridge_design}(a) and~(b), respectively. Waves ideally propagate inside the LC between the two metallic parallel plates that form the ridge and the upper plate. The phase shift is then electronically controlled by polarizing the LC and changing the orientation of the molecules. Similarly to the design recently proposed in~\cite{ridgegap_phaseshifter}, a container made of Rexolite is employed to confine the~LC and prevent its leakage. The bed of nails inserted at both sides of the liquid crystal acts as an artificial magnetic conductor (AMC), creating a high impedance surface condition. For computation purposes, an LC mixture GT3-23001 has been used: $\varepsilon_\bot=2.46$, $\Delta\varepsilon=0.82$, $\tan \delta_\bot=0.0143$, $\tan \delta_{\parallel}=0.0038$, and Rexolite of electrical parameters $\varepsilon_{r}^{\mathrm{Rexo}}=2.33$ and $\tan \delta^\mathrm{Rexo}=0.00066$ \cite{ridgegap_phaseshifter}.

\begin{figure}[!t]
	\centering
	\subfigure[]{\includegraphics[width= 0.93\columnwidth]{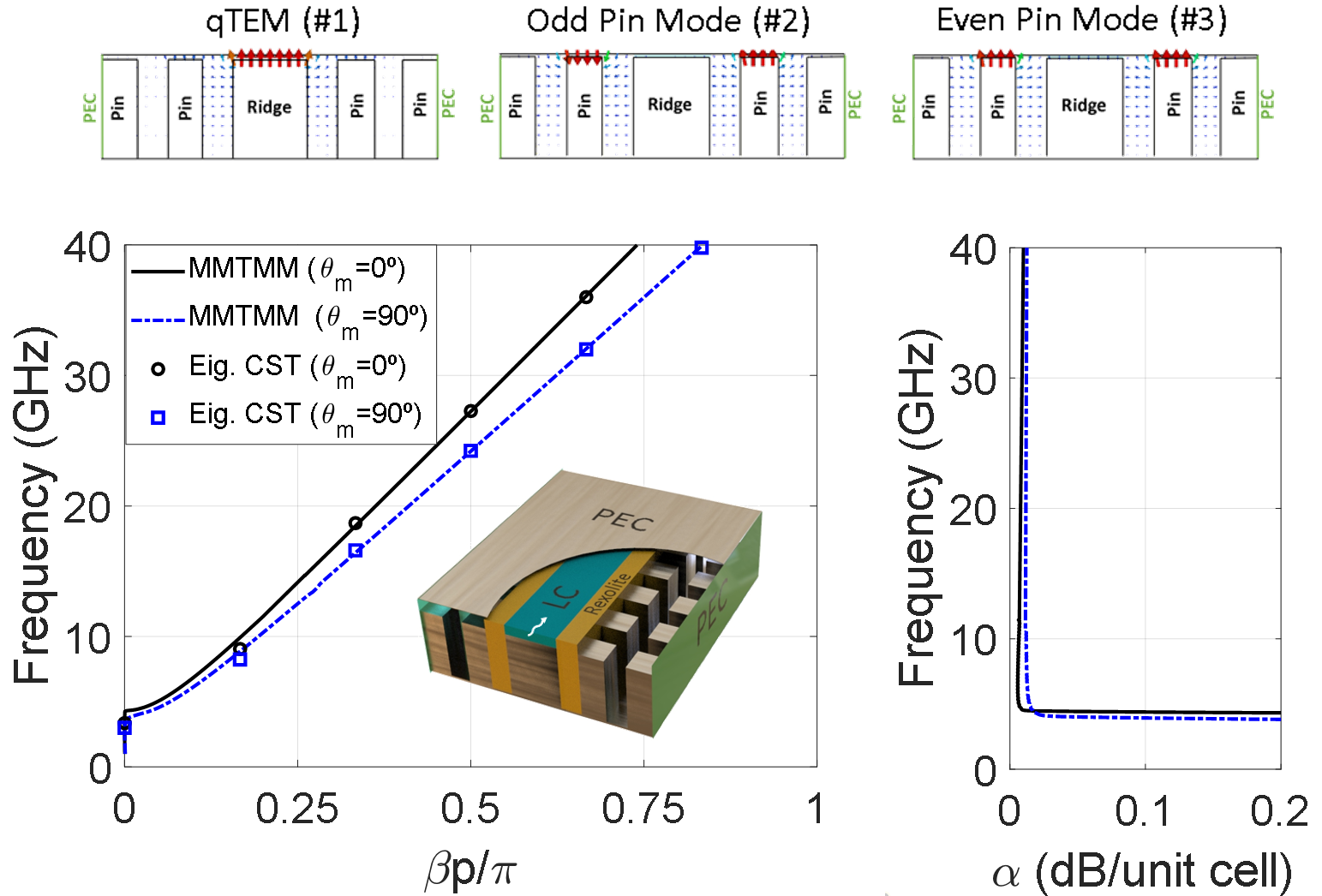}
	}
	\subfigure[]{\includegraphics[width= 0.93\columnwidth]{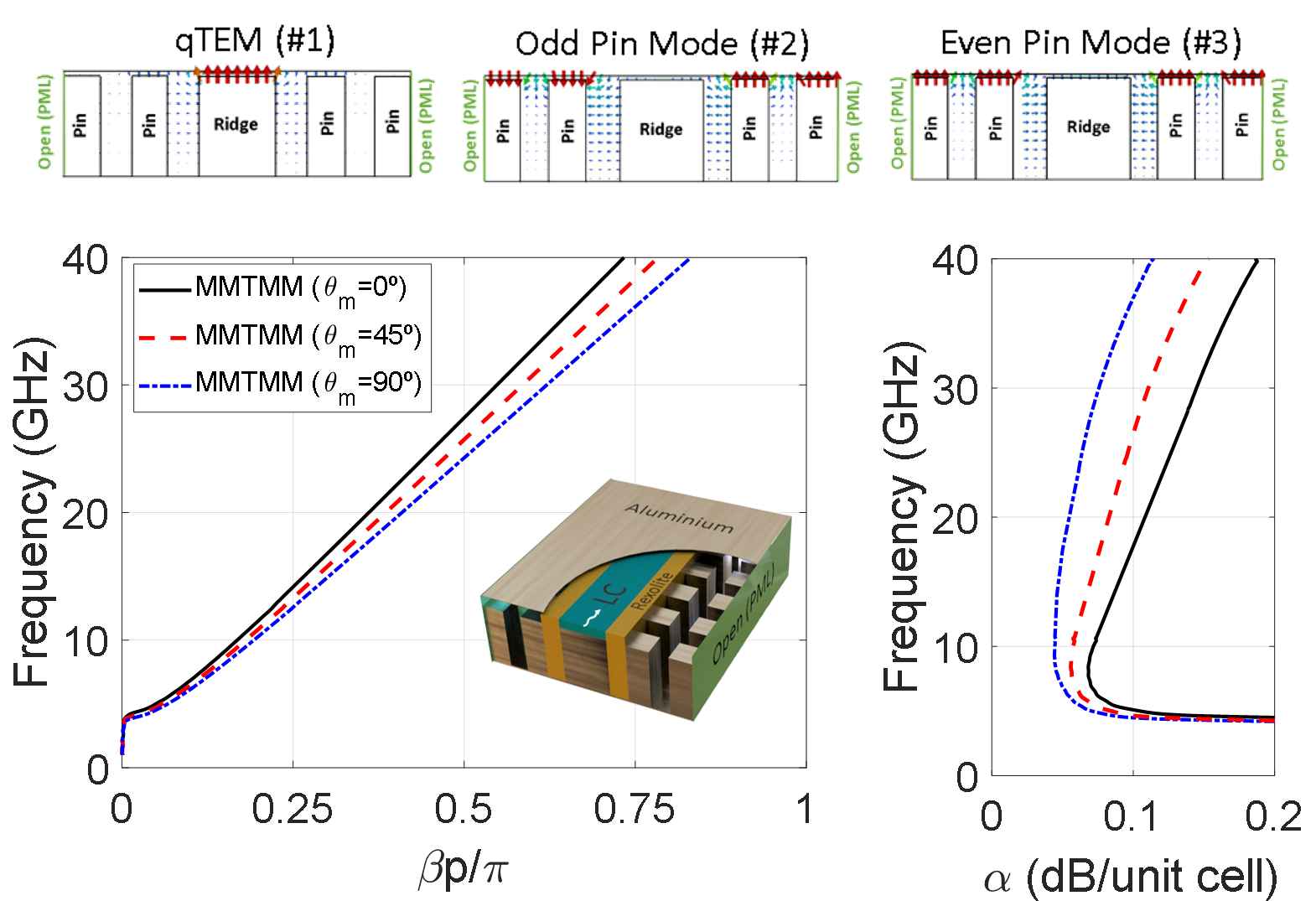}
	}
	\caption{Dispersion diagram of a (a)~lossless and (b)~lossy ridge gap waveguide phase shifter filled with liquid crystal. The results are obtained with $N=3$ modes. The electric field distribution of the three considered modes is displayed in the figure. The electrical and geometrical parameters of the unit cell are: $\varepsilon_\bot=2.46$, $\Delta\varepsilon=0.82$, $\tan \delta_\bot=0.0143$, $\tan \delta_{\parallel}=0.0038$, $\varepsilon_{r}^{\mathrm{Rexo}}=2.33$, $\tan \delta^\mathrm{Rexo}=0.00066$,   $p=1.76$\,mm, $w_\mathrm{LC}=2$\,mm, $h_\mathrm{LC}=80\, \mu$m, $w_p=0.96$\,mm, $h_p=2.64$\,mm, and $w_r=0.8$\,mm. } 
	\label{ridge_dispersion}
\end{figure}

Figs.\,\ref{ridge_dispersion}(a) and~(b) show the dispersion diagrams of the reconfigurable phase shifter in lossless and lossy scenarios, respectively, for different average tilt angles $\theta_m$. Both diagrams have been computed with $N=3$ modes in a time-domain simulation in \textit{CST}. No de-embeding layers are utilized in this case, since we are essentially working with diagonal tensors (cases $\theta_m=0,90^\mathrm{o}$) and the off-diagonal permittivity terms, $\varepsilon_{yz}=\varepsilon_{zy}$,  are negligible in this configuration for the case  $\theta_m=45^\mathrm{o}$. The electric field distributions of the three modes considered in the input/output ports are displayed in the figures. Note that modes~\#2 and~\#3 are included to take into account wave propagation along the pins  and ensure a correct convergence of the method.  The cutoff frequency of the propagating quasi-TEM mode, located approximately at~2.95\,GHz, is evidenced in both subfigures. Below the cutoff, even and odd quasi-TEM modes can propagate between the pins and the upper metal plate with a high attenuation constant~\cite{ridge_Kildal}.

In \Fig{ridge_dispersion}(a), the values of the phase and attenuation constants computed with the MMTMM in a lossless and bounded (PEC as lateral boundaries) structure are shown and compared with the data provided by the~\textit{CST} eigenmode solver. A good agreement is found between both set of results. Although the qTEM mode should have a null attenuation constant ($\alpha=0$) in the propagating frequency range (due to the absence of losses), small values of attenuation appear in the MMTMM due to inevitable numerical noise.

\Fig{ridge_dispersion}(b) presents a more realistic scenario, where the phase and attenuation constants are computed in a lossy and open (PML as lateral boundaries) structure. Material losses have been included in the liquid crystal and in the Rexolite, and PEC layers have been replaced by aluminium. As the \textit{CST} eigenmode solver does not provide the dispersion diagram in lossy and open structures, no comparison with~\textit{CST} appears in~\Fig{ridge_dispersion}(b). Three different values of $\theta_m$ have been considered: 0º, 45º and 90º. The structure becomes denser as $\theta_m$ increases as a consequence of the corresponding increment of the the effective permittivity of the LC. Also, it is observed a progressive increase of the attenuation constant as frequency increases, associated with the longer electrical length that the wave has to travel at higher frequencies. The case $\theta_m=0$ shows the highest attenuation constant, since for this configuration the effective term of the loss tangent tensor ($\tan \delta_{zz}=\tan \delta_{\bot}$) has the greatest value. 


\begin{figure}[!t]
	\centering
	\subfigure[]{\hspace*{-0.5cm}
\includegraphics[width= 0.95\columnwidth]{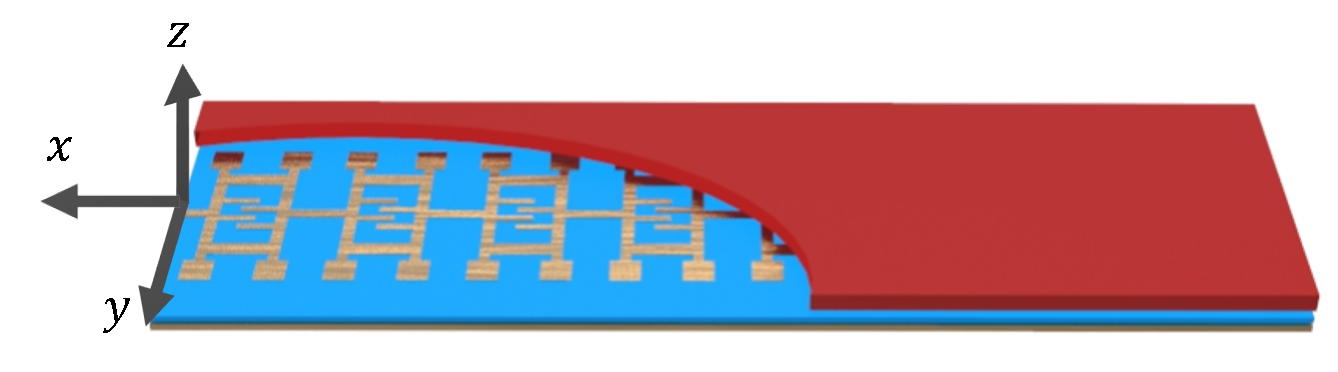}
	}\\
	\subfigure[]{\includegraphics[width= 0.60\columnwidth]{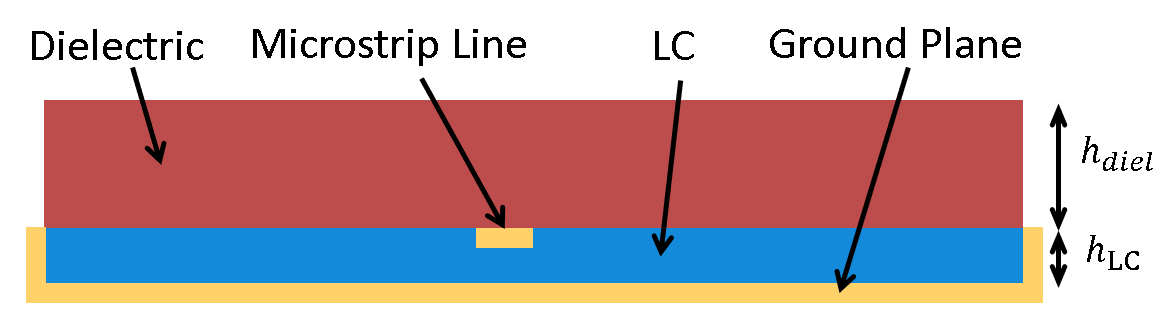}
	}%
	\subfigure[]{\includegraphics[width= 0.4\columnwidth]{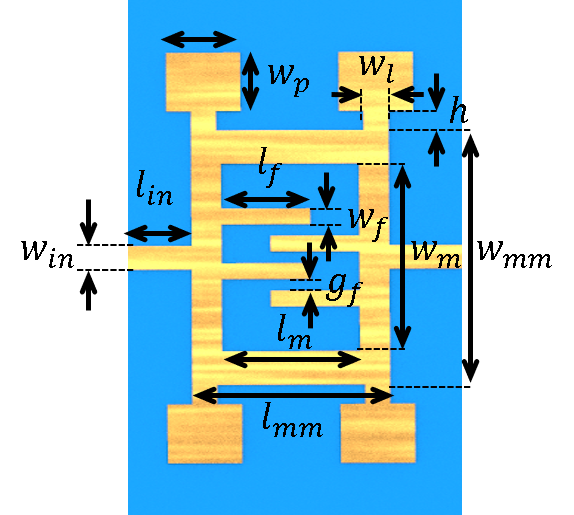}
	}\\
	\caption{(a) Liquid-crystal-based reconfigurable leaky-wave antenna presented in~\cite{lw_packaging}. (b)~Transversal cut view showing its forming layers. (c)~Unit cell.} 
	\label{leakywave}
\end{figure}

\section{\label{sec:LWA}   Radiating Structures: LC-based LWA}
 In this section, by following the recommendations given in Sec. II.C for open and radiating structures, we show that the MMTMM can be efficiently applied for the computation of the dispersion properties of  reconfigurable leaky-wave antennas (LWAs) based on the use of liquid crystal. For validation purposes, we replicate and analyze the design presented in~\cite{lw_packaging}, which is one of the few LC-based LWAs reported in the literature that has been manufactured and experimentally measured. \Fig{leakywave} shows the schematic of the LWA; namely, a composite right/left-handed (CRLH) LWA implemented in microstrip technology, the substrate of which is a LC to provide reconfigurability. The radiation angle of the LWA is controlled by polarizing the LC and, therefore, changing the average orientation of the molecules~($\theta_m$).  To confine the LC and prevent its leakage, there is a groove on the metal base forming a cavity [see~\Fig{leakywave}(b)]  in combination with a dielectric slab above the LC. The geometrical parameters of the LC-based LWA are given in Table\,\ref{tab:my-table}. The LC utilized here is TUD-649 ($\varepsilon_\bot=2.46$, $\Delta\varepsilon=0.82$) and the dielectric slab has a relative permittivity $\varepsilon_r^\mathrm{diel}=3.66$ \cite{lw_packaging}.

\begin{table}[!t]
\centering
\caption{GEOMETRICAL PARAMETERS OF THE LC-BASED LWA.}
\label{tab:my-table}
\begin{tabular}{|c|c|c|c|c|c|}
\hline
\textbf{Parameters} & \textbf{$l_m$}    & \textbf{$w_m$} & \textbf{$l_{mm}$} & \textbf{$w_{mm}$} & \textbf{$l_{in}$}   \\ \hline
Value (mm) & 4   & 6.5 & 2.8 & 4.8  & 1.4   \\ \hline
\textbf{Parameters} & \textbf{$h$}      & \textbf{$w_l$} & \textbf{$l_p$}    & \textbf{$w_p$}    & \textbf{$l_f$}      \\ \hline
Value (mm) & 0.5 & 0.5 & 1.5 & 1.5  & 1.8   \\ \hline
\textbf{Parameters} & \textbf{$w_{in}$} & \textbf{$w_f$} & \textbf{$g_f$}    & \textbf{$h_\text{LC}$} & \textbf{$h_\text{diel}$} \\ \hline
Value (mm) & 0.6 & 0.4 & 0.3 & 0.25 & 0.762 \\ \hline
\end{tabular}
\end{table}

\begin{figure}[t]
	\centering
	\subfigure[]{\includegraphics[width= 0.99\columnwidth]{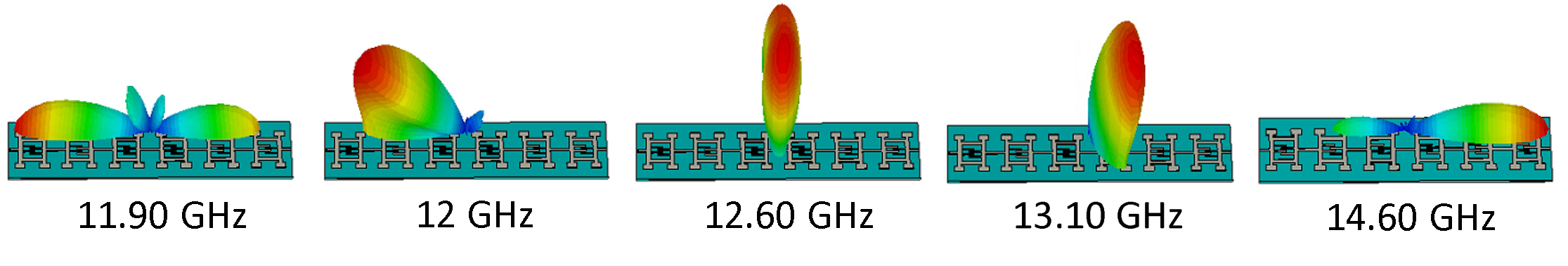}
	}
	\subfigure[]{\includegraphics[width= 0.99\columnwidth]{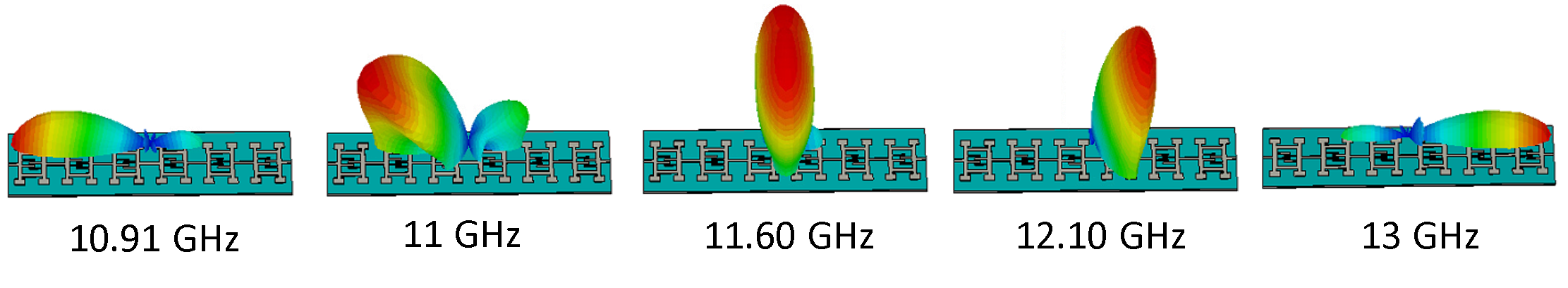}
	}
	\caption{3-D radiation pattern of the liquid-crystal-based reconfigurable leaky-wave antenna presented in~\cite{lw_packaging} for (a)~$V=0$ and (b)~$V\rightarrow \infty$ at different frequencies. Input and output ports are located on the left and right side of the structure, respectively.  } 
	\label{lw_3Dpattern}
\end{figure}
\subsection{Dispersion diagram and radiation properties}

\Fig{lw_3Dpattern} shows the 3-D radiation patterns of the LC-based LWA at different frequencies for the extreme polarization voltages $V=0$ ($\theta_m=0^\mathrm{o}$) and $V \rightarrow \infty$ ($\theta_m=90^\mathrm{o}$). For computational purposes, six unit cells are concatenated and the time-domain solver of \textit{CST} is used. As shown, the antenna is able to radiate at broadside direction [12.60\,GHz in \Fig{lw_3Dpattern}(a) and 11.60\,GHz in \Fig{lw_3Dpattern}(b)] due to the capacitance of the interdigitated structure placed at the center of the unit cell and the shunt inductance obtained by the stub lines connected to the square patches [see~\Fig{leakywave}(c)]. Below and above the broadside frequency, the antenna scans in backward (BW) and forward (FW) radiation angles.

\begin{figure}[t!]
	\centering
	\subfigure[]{\includegraphics[width= 0.9\columnwidth]{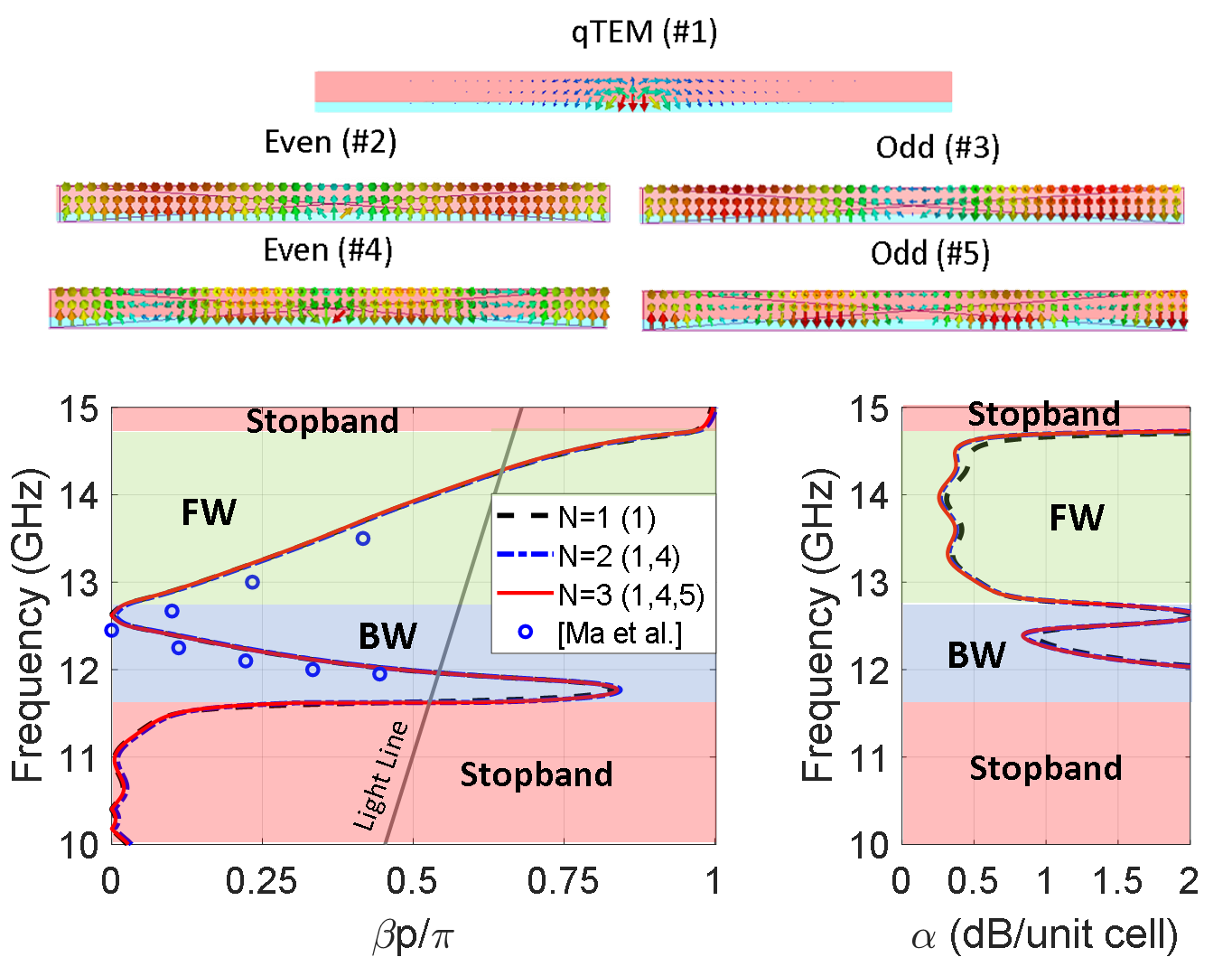}
	}
	\subfigure[]{\includegraphics[width= 0.92\columnwidth]{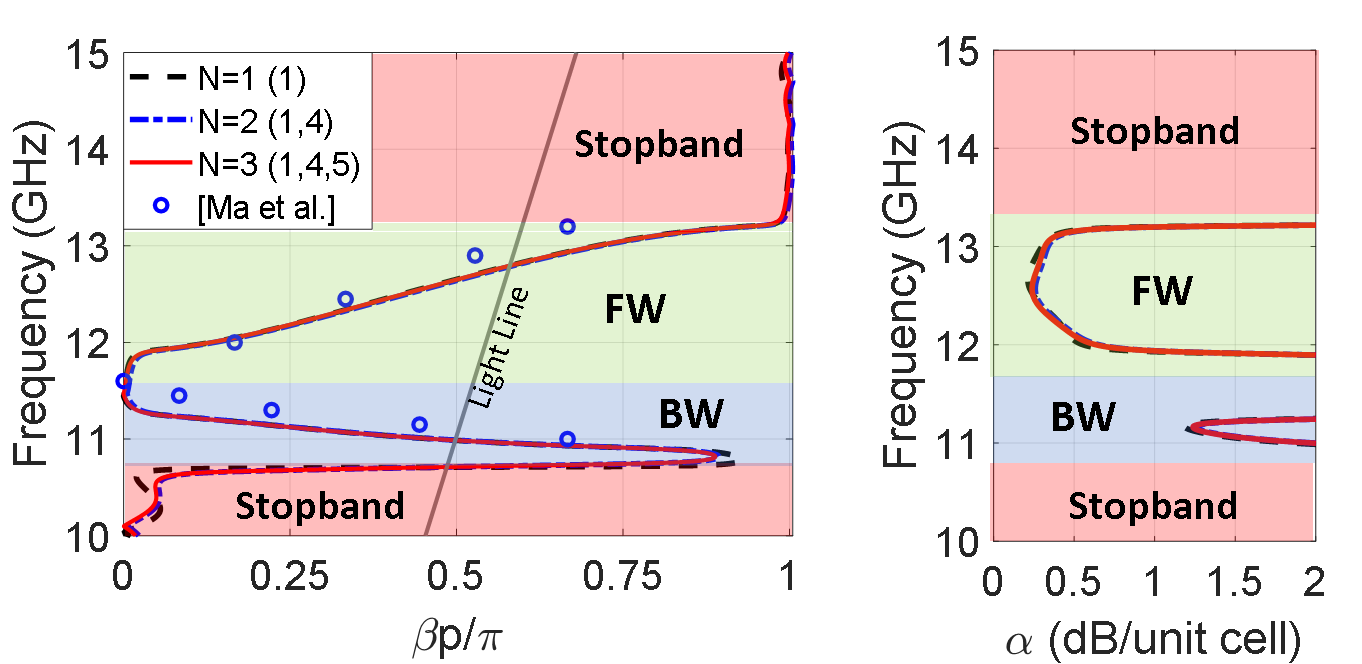}
	}
	\caption{Dispersion diagrams of the liquid-crystal-based reconfigurable leaky-wave antenna presented in \cite{lw_packaging} for (a) $V=0$ and (b) $V\rightarrow \infty$. The electrical parameters of the LC are: $\varepsilon_\bot=2.43$, $\Delta\varepsilon=0.79$. } 
	\label{lw_dispersion}
\end{figure}

The dispersion diagrams of the lossless LC-based LWA are shown in~Figs.\,\ref{lw_dispersion}(a) and~(b) when a null and a hypothetical infinite voltage is applied to the LC, respectively. The generalized scattering parameters of the unit cell have been extracted by performing a time-domain simulation in \textit{CST}. No de-embedding layers have been included in this case, since we are only considering the extreme polarization states of the LC, where the permittivity tensor is  diagonal.  The results extracted from the MMTMM are in good agreement with the data provided in~\cite{lw_packaging}.  Since the structure is not symmetric, it is expected that modes of even and odd parity are required to ensure the convergence of the attenuation and phase constants. In this case, two modes ($N=2$) are required to achieve that convergence. The combination of modes in the input/output ports that provides the best results includes the modes~\#1, \#4, and~\#5 depicted in the top panel of~\Fig{lw_dispersion}. The results in this figure show that mode~\#4 has relevance in the computation, due to its even nature and the high field intensity near the area of the microstrip line where the qTEM mode propagates. Modes~\#2 and~\#3 strongly depend on the size of the \textit{bounded} input/output  ports and, therefore, are hardly correlated to the Floquet modes that can physically propagate in the \textit{unbounded} periodic structure. 

Backward, forward, and stopband regions are shadowed in \Fig{lw_dispersion}.  In~\Fig{lw_dispersion}(a), the broadside frequency (corresponding to~$\beta=0$) is observed at~12.65\,GHz, which is in good agreement with the central frequency of the 3-D pattern displayed in~\Fig{lw_3Dpattern}(a). At~11.97\,GHz, the light line crosses the backward mode ($\beta=-k_0$). This frequency point is associated to backfire radiation in the antenna, which is also evidenced in~\Fig{lw_3Dpattern}(a) at~11.90\,GHz. Furthermore, note that the reduced slope of the phase constant in the backward region indicates that the scan angle rapidly changes in~\Fig{lw_3Dpattern}(a) from backfire to broadside radiation. On the contrary, the slope of the phase constant is higher in the forward region, which indicates a large scanning bandwidth. At~14.35\,GHz, the light line crosses the forward mode ($\beta=k_0$). This frequency point is associated with endfire radiation in the antenna, which can be observed at~14.60\,GHz in~\Fig{lw_3Dpattern}(a). In~\Fig{lw_dispersion}(b), backfire, broadside, and endfire frequencies are located at~11\,GHz, 11.60\,GHz, and~12.80\,GHz, respectively, in good agreement with the radiation patterns shown in~\Fig{lw_3Dpattern}(b).

\subsection{Optimization of the LWA via the Multi-modal Technique}


\begin{table}[!t]
\centering
\caption{GEOMETRICAL PARAMETERS OF THE OPTIMIZED LWA.}
\label{tab:optimized}
\begin{tabular}{|c|c|c|c|c|c|}
\hline
\textbf{Parameters} & \textbf{$l_m$}    & \textbf{$w_m$} & \textbf{$l_{mm}$} & \textbf{$w_{mm}$} & \textbf{$l_{in}$}   \\ \hline
Value (mm) & 4.5   & 6.5 & 2.9 & 5.4 & 0.75\\ \hline
\textbf{Parameters} & \textbf{$h$}      & \textbf{$w_l$} & \textbf{$l_p$}    & \textbf{$w_p$}    & \textbf{$l_f$}      \\ \hline
Value (mm) & 0.6 & 0.5 & 1.55 & 1.55  & 1.8   \\ \hline
\textbf{Parameters} & \textbf{$w_{in}$} & \textbf{$w_f$} & \textbf{$g_f$}    & \textbf{$h_\mathrm{LC}$} & \textbf{$h_\mathrm{diel}$} \\ \hline
Value (mm) & 0.55 & 0.4 & 0.3 & 0.25 & 0.762 \\ \hline
\end{tabular}
\end{table}

\begin{figure}[t!]
	\centering
	\subfigure[]{\includegraphics[width= 0.9\columnwidth]{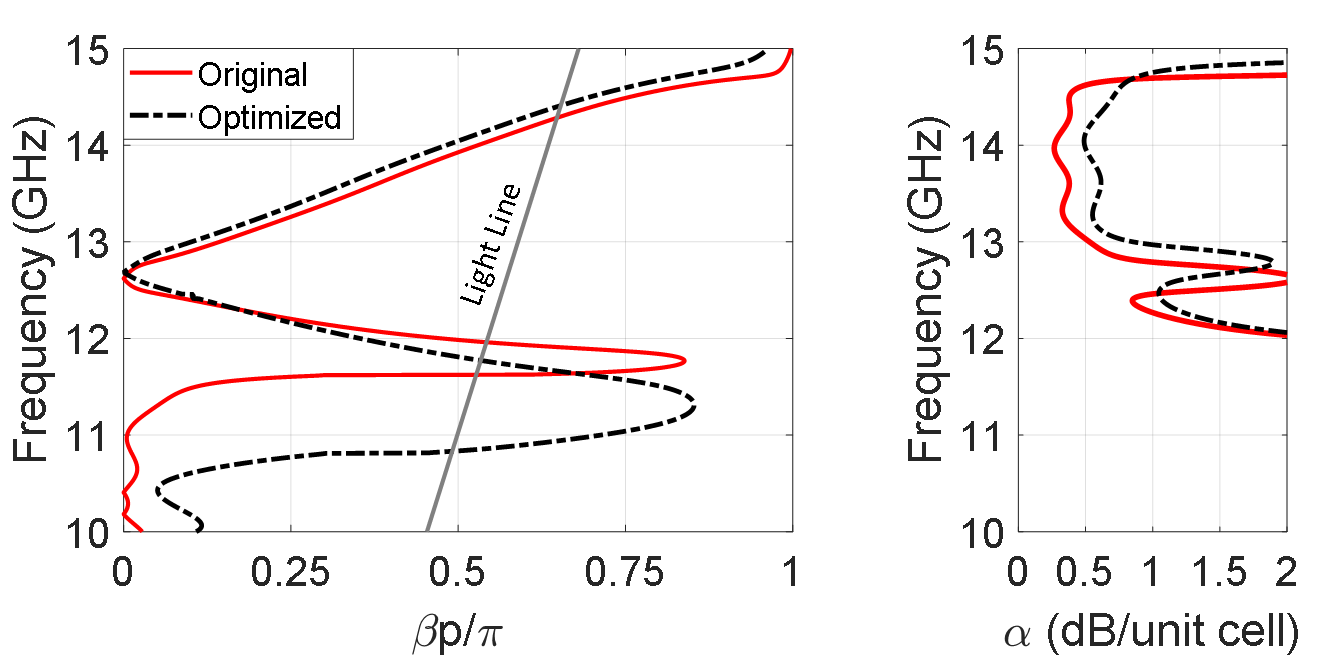}
	}
	\subfigure[]{\includegraphics[width= 0.85\columnwidth]{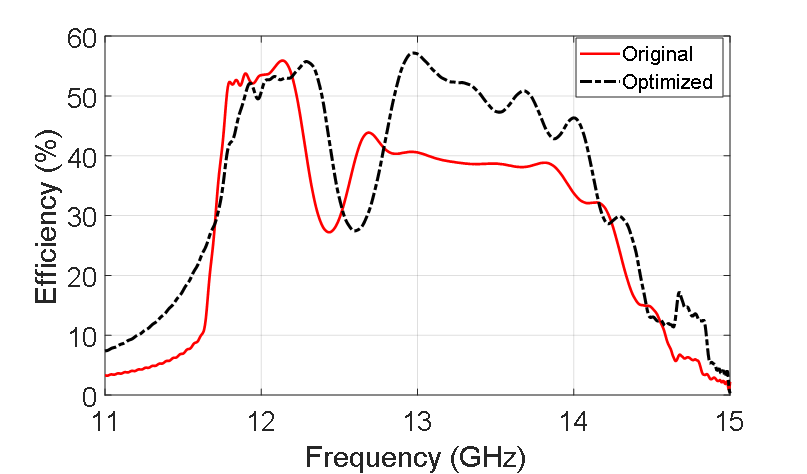}
	}
	\caption{(a) Dispersion diagram  of the optimized LC-based  leaky-wave antenna for $V=0$. (b) Radiation efficiency of the optimized antenna for $V=0$ when 8 unit cells are cascaded. The electrical parameters of the LC are: $\varepsilon_\bot=2.43$, $\Delta\varepsilon=0.79$. } 
	\label{lwOpt}
\end{figure}

Up to this point, the MMTMM has been used only as an \textit{analysis tool}, that is, to extract the dispersion properties of the structures under consideration. Next, we will show that the MMTMM has also potential to be used as a \textit{design tool}; in this case, to increase the leakage rate and, therefore, the radiation efficiency of the previously analyzed leaky-wave antenna. The results of this study led us to the geometrical parameters of the optimized leaky-wave antenna shown in~Table\,\ref{tab:optimized}. To make a fair comparison, the original shape of the unit cell and thicknesses of the LC and dielectric have been preserved; that is, no additional microstrip sections have been added to the unit cell.  The dispersion diagram of the optimized LWA is shown in~\Fig{lwOpt}(a). It can be observed in that figure that the phase constant of the optimized antenna shows a similar behavior as the original antenna and, consequently,  the radiation angles of both antennas are similar. However, the leakage rate $\alpha$ has been enhanced considerably, specially in the forward region, which directly translates in a higher efficiency of the optimized antenna. This fact is evidenced in~\Fig{lwOpt}(b), where the radiation efficiencies of the original and optimized antennas are compared. The radiation efficiency is improved an 11\% in the forward region and kept similar in the backward region. The drop in the efficiency observed around~12.5\,GHz in~\Fig{lwOpt}(b) for both antennas can be related to the appearance of the so-called open stopbands~\cite{Paulotto2009, Williams2013}. This stopband appears in many periodic LWAs when the beam is scanned through broadside and gives rise to peaks in the attenuation constant, as the ones appearing in~\Fig{lwOpt}(a) around~12.7\,GHz. The frequency shift found between the efficiency drop in~\Fig{lwOpt}(b) and the attenuation peaks in~\Fig{lwOpt}(a) is associated with the finite size (eight cells) of the periodic LWA analyzed in~\Fig{lwOpt}(b) versus the infinite nature of the periodic structure considered in the dispersion diagram of~\Fig{lwOpt}(a).  A parametric study revealed that the open stopband, in agreement with the rationale reported in~\cite{Paulotto2009, Williams2013}, cannot be easily suppressed with the current configuration of the unit cell. A modified configuration of the antenna would be needed, which could be conveniently analyzed by means of the proposed MMTMM.


\section{\label{sec:Conc} Conclusion}
The use of the multi-modal transfer-matrix method to compute the dispersion diagram of periodic structures involving general anisotropic media has been discussed in this work. We have particularized the study to the case of liquid crystals due to its promising properties for the design of electronically reconfigurable devices. The proposed method, which combines the use of commercial simulators and analytical post-processing, overcomes the common limitations of general commercial eigenmode solvers when dealing with anisotropic materials. Specifically, the proposed multi-modal method shows three interesting properties:
\begin{enumerate}
    \item Anisotropic materials with non-diagonal permittivity and permeability tensors can be analyzed. This is an interesting feature that has been exploited throughout the text, since commercial eigenmode solvers find difficulties  when computing the complex propagation constant in some configurations involving non-diagonal tensorial materials.
    \item The attenuation constant can be easily computed. This is of capital relevance in order to analyze the stopband regions of periodic structures. Furthermore, the method allows to include lossy materials in the computation. This is a very appreciated feature in LC-based structures in order to take into account the lossy nature of the material. 
    \item Unbounded and radiating structures can be analyzed. From the aforementioned statements, this is the most remarkable feature of the proposed method. Conversely, in most of commercial eigenmode solvers the structure must be forcefully shielded with perfect electric/magnetic boundary conditions. This fact prevents that periodic leaky wave antennas can be typically analyzed in commercial software.
\end{enumerate}

Some relevant works in the literature were selected to test the method. We started with the study of canonical waveguide and microtrip sections. Afterwards, we apply the multi-modal method to analyze the dispersion properties (phase shift, radiation angle, leaky rate, etc.) of more advanced designs, such as a reconfigurable phase shifter in ridge gap-waveguide technology and a leaky-wave antenna. All results were in good agreement with previously reported works and with commercial software \textit{CST} and \textit{Ansys \textit{HFSS}}, demonstrating  that the multi-modal  method  has potential application in the analysis and design of  periodic and uniform structures  that includes liquid crystal or other generic anisotropic materials.

\section*{Acknowledgment}
The authors would like to thank Prof. Armando Fernández-Prieto for his valuable help.

\ifCLASSOPTIONcaptionsoff
  \newpage
\fi

\end{document}